\documentclass[aps,pre,reprint,nofootinbib]{revtex4-1}

\usepackage{amsmath}
\usepackage{amssymb}
\usepackage{amsfonts}
\usepackage{bm}
\usepackage{graphicx}

\newcommand{\avg}[1]{\langle#1\rangle} 
\newcommand{\prm}[1]{#1^\prime} 
\newcommand{\oper}[1]{\mathcal{#1}} 
\newcommand{\mtx}[1]{\bm{#1}} 
\newcommand{\ceil}[1]{\lceil{#1}\rceil} 
\DeclareMathOperator{\std}{std} 

\begin{document}
\title{
  Fluctuation theory in space and time:\\
  white noise in reaction-diffusion models of morphogenesis
}
\date{\today; Revision: 1.0}
\author{Roman Belousov}\email{belousov.roman@gmail.com}
\author{Adrian Jacobo}
\author{A. J. Hudspeth}
\affiliation{Howard Hughes Medical Institute, Laboratory of Sensory Neuroscience, The Rockefeller University New York, NY 10065, USA}
\begin{abstract}
  The precision of reaction-diffusion models for mesoscopic physical systems is limited
  by fluctuations. To account for this uncertainty, Van Kampen derived a stochastic
  Langevin-like reaction-diffusion equation that incorporates spatio-temporal white
  noise. The resulting solutions, however, have infinite standard deviation. {\it Ad hoc}
  modifications that address this issue by introducing microscopic correlations are
  inconvenient in many physical contexts of wide interest. We instead estimate the
  magnitude of fluctuations by coarse-graining solutions of the Van Kampen equation
  at a relevant mesoscopic scale. The ensuing theory yields fluctuations of finite
  magnitude. Our approach is demonstrated for a specific biophysical model---the
  encoding of positional information. We discuss the properties of the fluctuations
  and the role played by the macroscopic parameters of the underlying reaction-diffusion
  model. The analysis and numerical methods developed here can be applied in physical
  problems to predict the magnitude of fluctuations. This general approach can also
  be extended to other classes of dynamical systems that are described by partial
  differential equations.
\end{abstract}
\maketitle

\section{\label{sec:intro}Introduction}
In addition to applications in chemistry and other disciplines \cite{Nicolis2007},
reaction-diffusion (RD) equations are commonly accepted as the basis of morphogenetic
models in biology \cite{Turing1952,Wolpert1969,*Wolpert2016,Green2015,Tkacik2015,HalatekFrey2018}.
A classical example is the encoding of positional information (PI). During embryological
development, an organism must be partitioned into distinct morphological and functional
components. The positions of these structures may be specified by a chemical agent---a
{\it morphogen}---whose local concentration varies across the embryo and obeys RD
equations. In this context one encounters perhaps the simplest example of such systems,
which has been chosen to demonstrate the theory presented in this paper.

\begin{figure}
\includegraphics[width=1\columnwidth]{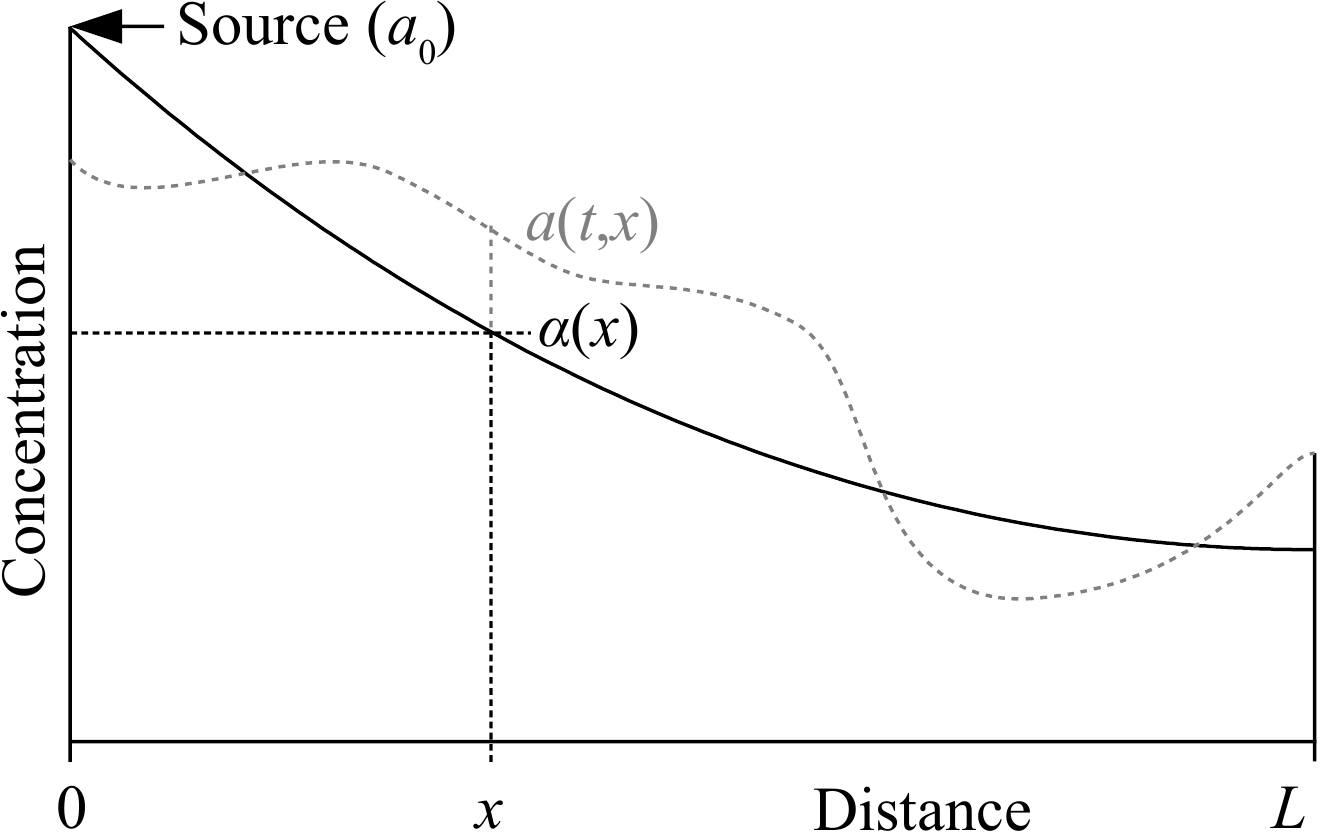}
\caption{\label{fig:xmp}
  A morphogen produced at the left boundary propagates by diffusion into the rest
  of the one-dimensional system $\Lambda=[0,L]$. The morphogen's steady-state concentration
  $\alpha(x)$ owing to a degradation reaction decreases monotonically towards the
  impenetrable right boundary. Each position coordinate $x\in\Lambda$ corresponds
  to a unique value of the positional-information curve $\alpha(x)$. The instantaneous
  concentration of the morphogen $a(t,x)$, however, is subject to spontaneous fluctuations
  on a mesoscopic scale.
}
\end{figure}

As a typical RD system we consider the dynamics of a single morphogen that diffuses
from a localized source over a confined spatial domain and undergoes chemical degradation
(Fig.~\ref{fig:xmp}). Once all transients have decayed and the system has reached
a steady state, cells or organelles can measure their distance to the source by reading
out the local concentration of the morphogen. For this reason it is said that the
morphogen encodes PI.

Quantitative characterization of PI noise is an important problem in biophysics
\cite{Gregor2007Probing,Gregor2007Stability,Bollenbach2008,Dubuis2013,Tkacik2015,Buceta2017}.
Because many key processes in development occur on micrometer scales, the underlying
chemical reactions and diffusive flows are subject to spontaneous variations. These
fluctuations disrupt the local concentration of the morphogen and reduce the amount
of information that a RD system contains \cite{Gregor2007Probing,Dubuis2013,Tkacik2015}.
A relevant question is then: how reliably can PI be encoded and read out in the presence
of noise?

Most studies concentrate on the problem of decoding PI. For example, one can estimate
the efficiency with which a cell measures a morphogen's concentration \cite{BergPurcell1977,Gregor2007Probing}.
The concentration can also be measured \cite{Gregor2007Probing,Gregor2007Stability};
the experimental precision then provides an upper bound for the uncertainty of PI.
Our understanding of the readout problem is incomplete, however, for one should also
take into account how much information a noisy RD system actually contains.

The physical theory of fluctuations opens an avenue to the problem of encoding PI.
At the mesoscopic scale, the dynamics of a reaction-diffusion system can be described
by a stochastic partial differential equation derived from simplified microscopic
mechanics \cite{vanKampenXIV,vanKampen1976,Gardiner8}. The noise level in this model
is completely determined by the macroscopic parameters of the system, such as the
diffusion and reaction constants. The magnitude of fluctuations in the morphogen's
concentration should then in theory be calculable. This approach promises clearer
results on the precision of PI than the analysis of empirical data.

As shown in Sec.~\ref{sec:thr}, the Van Kampen equation leads to a solution of infinite
variance and therefore also of infinite standard deviation. Because both of these
statistics measure the magnitude of fluctuations, one may regard this result as futile
and seek a more realistic model. The existing alternatives
\cite{[{Chapter 2 in }] []GOjalvaSancho,[{Chapter 8 in }] []Kotelenez,[{Chapter 1 in }] []Holder,[{Sec. 1.2.5 in }] []Lototsky}
either part ways with the Van Kampen equation or require an additional, {\it ad hoc}
layer of theory. Both approaches, however, rely on new phenomenological constants
such as the amplitude or correlation length of microscopic noise. Although these
parameters control and regulate the fluctuation's magnitude, they can be inferred
neither from the meso- or macroscopic dynamics nor from the ensuing theory itself.
Because one can only fit the new parameters to observations, these models are purely
descriptive. This lack of predictive power is one reason why the theoretical avenue
to the problem of encoding PI has received little attention.

In contradistinction to the theoretical approaches mentioned above, we estimate the
fluctuations in a PI problem by solving the Van Kampen equation without modifications.
A plausible level of noise is obtained if the resultant morphogen concentration is
integrated in space over a subscale of the RD system. This procedure is consistent
with classical fluid dynamics, in which macroscopic fields are commonly understood
as coarse-grained representations of microscopic systems \cite{[{Sec. I.1 in }] []LandauLifshitz6,[{Sec. 1.2, p.6 in }] []KardarII}.

Multiscale models of computational physics, which combine the methods of finite elements
and molecular dynamics \cite{Mortazavi2013,Kojic2014}, make the coarse-graining procedure
even more explicit. The macroscopic properties of a molecular-dynamics system are
calculated as spatial averages by means of a microscopic connection \cite[Sec. 3 and 6]{EvansMorriss}.
The exact procedure amounts to integration of molecular degrees of freedom over volume,
which is suggestive of the coarse-graining subscale. The finite-element method then
offers techniques to solve the dynamical equations of macroscopic fields. Note that,
as the volume of a molecular-dynamics system decreases, the uncertainty of spatial
averages diverges, exactly as in the Van Kampen theory.

A coarse-graining subscale arises quite naturally in developmental biology: morphogenetic
features are not point-like, but have a finite mesoscopic extent in space. Moreover,
developmental decisions are often delegated to whole cells or to large organelles
such as cellular nuclei \cite{Gregor2007Probing,Buceta2017}. In the RD problems of
morphogenesis and PI, one should therefore reckon with the total amount of the substance
and its fluctuations over the scale of the target biological structure, rather than
with the concentration field at isolated points.

In the next section we briefly describe the Van Kampen equation and the coarse-graining
of its solution for a simple RD system. The implications of this theory and some
numerical results are then discussed in Sec.~\ref{sec:num}. Additional mathematical
details are provided in the Appendices. In particular, Appendices~\ref{sec:SM} and
\ref{sec:CM} concern two classes of the finite-element method used to simulate numerically
the dynamics of the coarse-grained stochastic fields.

\section{\label{sec:thr}Theory}
Because RD problems in general may not yield to analytical techniques, we use as
a case study our earlier example of a simple one-dimensional system (Fig.~\ref{fig:xmp}).
The associated theory can be treated by a variety of methods. Purely numerical techniques
then can be compared with a more accurate analytical approach. This example is not
entirely abstract, for it provides a model of the actual mechanism of PI encoding
in {\it Drosophila} embryos \cite{Gregor2007Probing,Grimm2010}.

In one dimension the number density of a morphogen $a(t,x)$, which depends on time
$t$ and position $x$, obeys the Van Kampen dynamic equation \cite{vanKampenXIV,vanKampen1976,Gardiner8}
\begin{equation}\label{eq:SRD}
  (\partial_t + k - D\partial_x^2) a(t,x) = f(t,x),
\end{equation}
in which the degradation rate $k$ and the diffusivity $D$ are positive constants,
whereas $f(t,x)$ represents microscopic noise. Appendix~\ref{sec:SRD} offers a short
justification of the Van Kampen equation.

The left-hand side of Eq.~(\ref{eq:SRD}) expresses the difference between the local
change of concentration $\partial_t a(t,x)$ and the classical nonequilibrium forces
of mass action and Fick's diffusion. In small systems the residual force
$f(t,x)$ does not vanish, but varies spontaneously because of microscopic events:
this is the origin of microscopic noise. To be consistent with classical fluid dynamics,
the steady-state ensemble averages of $f(t,x)$ and $a(t,x)$ must yield
\begin{equation}\label{eq:avgs}
  \avg{f(t,x)}=0,\quad\avg{a(t,x)} = \alpha(x),
\end{equation}
in which $\alpha(x)$---the PI curve---is the time-independent solution of the macroscopic
RD problem [Fig.~\ref{fig:xmp}; Eq.~(\ref{eq:alpha}) in Appendix~\ref{sec:SRD}].

A convenient model of the morphogen's source is a fixed-value condition imposed at
the left end of the interval $\Lambda=[0,L]$. In the macroscopic RD problem, this
constraint is supplemented quite naturally by a reflective right boundary, leading
to the expression~(\ref{eq:alpha}) for the concentration curve $\alpha(x)$. Nonetheless,
in Appendices~\ref{sec:SRD} and \ref{sec:bnd} we employ a different choice of the
right boundary condition for the stochastic equation~(\ref{eq:SRD}):
\begin{equation}\label{eq:bnd}
   a(t,x)\Big{|}_{x=0} = a_0 = \alpha(0),\quad a(t,x)\Big{|}_{x=L} = \alpha(L),
\end{equation}
with $a_0$ representing a source of constant strength. By virtue of Eq.~(\ref{eq:avgs}),
the fixed-value boundary condition remains macroscopically consistent with the reflective
boundary for the PI curve:
\begin{equation}\partial_x \alpha(x)\Big{|}_{x=L} = 0.\end{equation}
The problem of boundary conditions is addressed in Appendix~\ref{sec:bnd}.

At the mesoscale and in the tradition of fluctuation theory, Van Kampen relates $f(t,x)$
to two stochastic terms, owing to the fluctuations of the mass-action law and the
diffusive flow, respectively:
\begin{multline}\label{eq:ftx}
  f(t,x) = \sqrt{k \alpha(x)} \partial_x \dot{W}_1(t,x) \\+ \partial_x [\sqrt{2 D \alpha(x)} \partial_x \dot{W}_2(t,x)].
\end{multline}
Here $\partial_x \dot{W}_1$ and $\partial_x \dot{W}_2$ are independent, spatially
distributed, Gaussian white-noise variates of zero mean and unit strength (Appendix~\ref{sec:SRD}).
The overscript dots indicate the time derivatives. These noise sources are delta-correlated
in both space and time:
\begin{align}\label{eq:uncorrelated}
  &\avg{\partial_x \dot{W}_1\Big{|}_{t_1,x_1} \partial_x \dot{W}_2\Big{|}_{t_2,x_2}} = 0,
  \\\label{eq:white}
  &\avg{\partial_x \dot{W}_i\Big{|}_{t_1,x_1} \partial_x \dot{W}_i\Big{|}_{t_2,x_2}} = \delta(t_1-t_2) \delta(x_1-x_2),
\end{align}
which hold for $i=1,2$, with $\delta(\cdot)$ being the Dirac delta function. Note
in the above equations that spatially distributed white noise is singular in time
and space: its variance diverges as a product of two delta functions, $\lim_{t\to0}\delta(t)$
and $\lim_{x\to0} \delta(x)$.

To avoid immaterial details, we focus on the steady-state solution of Eq.~(\ref{eq:SRD}),
$a(\infty,x)$. We denote the deviation of the morphogen's concentration from the
ensemble average value by $\Delta{a}(t,x) = a(t,x) - \alpha(x)$. Then Eq.~(\ref{eq:steady})
of Appendix~\ref{sec:AM} gives us
\begin{equation}\label{eq:delta}
  \Delta{a}(\infty,x) = \lim_{t\to\infty}\int_0^t \prm{dt} \int_\Lambda \prm{dx} g(t-\prm{t},x|\prm{x}) f(\prm{t},\prm{x}).
\end{equation}
Here $g(t-\prm{t},x|\prm{x})$ is the Green's function that propagates disturbances
of the number density in time and space, from an instant $\prm{t}$ and position $\prm{x}$
to any other $t$ and $x$.

The steady-state variance of the deviation $\Delta{a}(\infty,x)$, as can be formally
calculated from Eq.~(\ref{eq:delta}), diverges [see also Eq.~(\ref{eq:variance})
in Appendix~\ref{sec:AM}]. To understand why this happens, apply the differential
chain rule to the second term on the right-hand side of Eq.~(\ref{eq:ftx}) and substitute
it into Eq.~(\ref{eq:delta}); one then finds the following term in the expression
for $\Delta{a}(\infty,x)$:
\begin{multline}\label{eq:divergent}
  \lim_{t\to\infty}\int_0^t \prm{dt} \int_\Lambda \prm{dx} g(t-\prm{t},x|\prm{x}) \sqrt{2 D \alpha(\prm{x})}
    \partial_{\prm{x}}^2 \dot{W}_2(\prm{t},\prm{x})
    \\ \sim \partial_{x} W_2(t,x).
\end{multline}
Here the time integration removes the temporal singularity of $\partial_x^2 \dot{W}_2$,
but the spatial integral is canceled by one of the two derivative operators. The
above term contains a spatial singularity of the order $\partial_{x} W_2(t,x)$
[Eq.~(\ref{eq:white})]. Therefore the variance of $\Delta{a}(\infty,x)$, expressed
formally by Eq.~(\ref{eq:variance}), in effect diverges.

If one replaces the positional delta function $\delta(\cdot)$ in Eq.~(\ref{eq:white})
by some bounded correlation kernel $C(\cdot)$ \cite[Sec. 2.1.2]{GOjalvaSancho}, the
spatial singularity disappears from Eqs.~(\ref{eq:delta}) and (\ref{eq:divergent}).
The cost of this approach is a significantly more complicated theory \cite{Stefanou2009,Sepahvand2010,[{Chapter 2 in }] []Ghanem}.
First, spatial noise correlations that regularize the variance of $\Delta{a}(\infty,x)$
must be modeled explicitly. Second, a nontrivial kernel $C(\cdot)$ increases the
mathematical difficulty of the problem. The spatial singularity of Eq.~(\ref{eq:divergent})
can alternatively be removed by integrating it with respect to the coordinate $x$,
the approach that we pursue here. An integration with respect to position occurs
when we coarse-grain the number density $a(t,x)$ over a scale $\xi$ of the appropriate
spatial dimension. Then, instead of the morphogen's concentration at some point $x$,
the quantity of interest becomes the {\it total number} of molecules in the $\xi$-neighborhood
$\Xi(x) = (x-\xi/2,x+\xi/2)$ of that point. If we use the inverse scale $\xi$ as
a normalization factor, we can equivalently consider a coarse-grained concentration
field
\begin{equation}
  a_\xi(x) = \int_{x-\xi/2}^{x+\xi/2} \frac{\prm{dx}}{\xi} a(\infty,\prm{x}).
\end{equation}

\begin{figure*}[!ht]
\includegraphics[width=2\columnwidth]{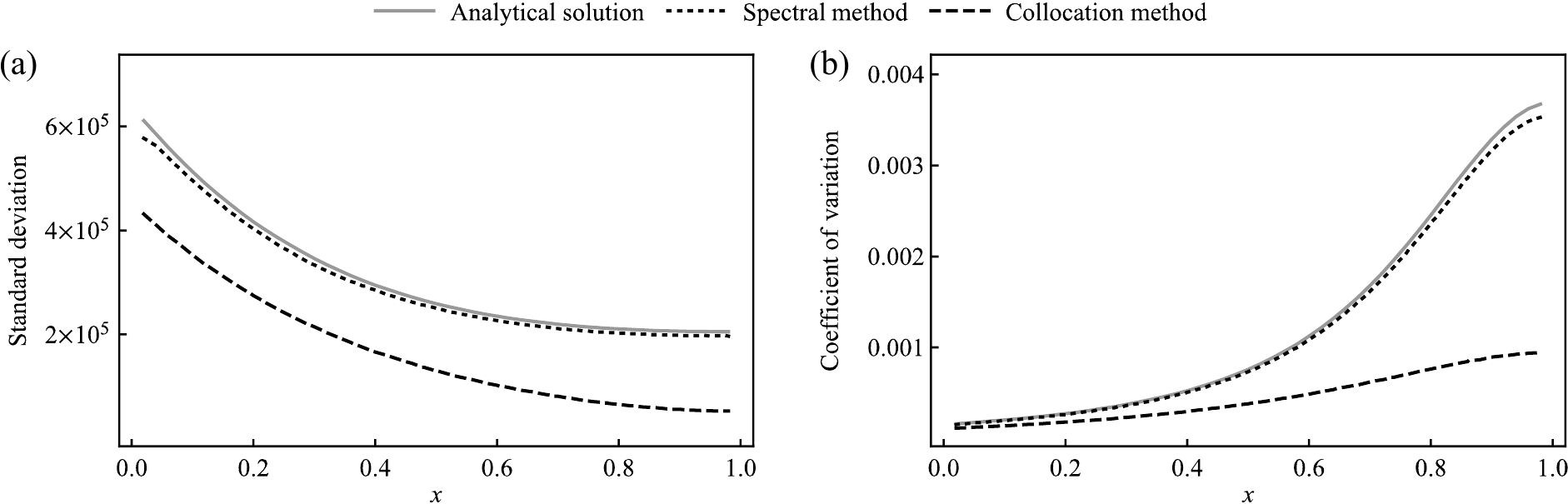}
\caption{\label{fig:num}
  Uncertainty of the coarse-grained concentration $\alpha_\xi(x)$ as a function of
  position for the experimental data of Refs.~\cite{Gregor2007Probing,Grimm2010}.
  The values of the standard deviation in A and the coefficient of variation in B
  are calculated by three methods: i) analytical solution (Appendix~\ref{sec:AM});
  ii) spectral finite-element simulations (Appendix~\ref{sec:SM}); iii) collocation
  finite-element method (Appendix~\ref{sec:CM}). The uncertainty of the PI curve
  does not exceed $0.4\,\%$ at any position.
}
\end{figure*}

The coarse-grained concentration of the morphogen undergoes fluctuations of finite
magnitude. As a statistical measure of this magnitude one can take either the variance
of $a_\xi$---the second cumulant $\kappa_2(a_\xi)$---or the standard deviation $\std(a_\xi)$.
By using the properties of the Green's and delta functions together with Eqs.~(\ref{eq:bnd})--(\ref{eq:delta}),
we find
\begin{multline}\label{eq:kappa2}
  \kappa_2[a_\xi(x)] =  \iint_{\Xi(x)} \frac{dx_1 dx_2}{\xi^2}
    \lim_{t\to\infty} \int_0^t \prm{dt}\int_\Lambda \prm{dx} \alpha(\prm{x})
    \\\times
    \Big{[}
      k g(t-\prm{t},x_1|\prm{x}) g(t-\prm{t},x_2|\prm{x})\\
      + 2 D \partial_{\prm{x}} g(t-\prm{t},x_1|\prm{x}) \partial_{\prm{x}} g(t-\prm{t},x_2|\prm{x})
    \Big{]}.
\end{multline}

A Fourier series expansion of the above expression, as well as of the coarse-grained
steady-state concentration $\alpha_\xi(x)$, is derived in Appendix~\ref{sec:AM} [Eqs.~(\ref{eq:smn})
and (\ref{eq:axi})]. The number density $\alpha_\xi(x)$ differs from $\alpha(x)$
by a factor that is negligible for small scales $\xi$. Both these fields interchangeably
represent a PI curve, for they convey nearly the same value everywhere in $\Lambda$.

A useful way to quantify the uncertainty of $a_\xi(x)$ is the coefficient of variation,
$\std(a_\xi)/\alpha_\xi$, which relates the level of fluctuations to the strength
of the PI signal. Quite generally, however, both the mean value of the morphogen's
concentration and its variance are proportional to the parameter $a_0$ [Eq.~(\ref{eq:smn})
and (\ref{eq:axi}), Appendix~\ref{sec:AM}]. Therefore the relative uncertainty is
inversely proportional to $\sqrt{a_0}$:
\begin{equation}\label{eq:sigma}
  \frac{\std[a_\xi(x)]}{\alpha_\xi(x)} = \frac{\sigma_\xi(x|\ell)}{\sqrt{a_0}},
\end{equation}
in which $\sigma_\xi(x|\ell)$ depends only on the coordinate $x$ and the parameters
$\xi$ and $\ell = \sqrt{D/k}L^{-1}$.

Fluctuations of physical quantities usually decay as the inverse square root of the
number of molecules involved \cite{[{Sec. I.2 in }] []LandauLifshitz5}. This dependence
is explicitly controlled by the parameter $a_0$ in Eq.~(\ref{eq:sigma}). The source
strength $a_0$ in the above expression should therefore be measured in one dimension
as the number of molecules per unit length. If molar or mass-density units are used
instead, Eq.~(\ref{eq:sigma}) does not render the coefficient of variation correctly.

Equation~(\ref{eq:sigma}) defines $\sigma_\xi(x|\ell)$, which we call a {\it variation
profile}. Given the values of $\xi$ and $\ell$, this relation can be evaluated numerically
as a function of position by use of Eqs.~(\ref{eq:smn}) and (\ref{eq:axi}). For a
source of any given strength $a_0$, the coefficient of variation---a rescaled variation
profile---can then be calculated easily from Eq.~(\ref{eq:sigma}).

Finally, the constant $\ell>0$ is the correlation parameter of $a_\xi(x)$ [Appendix~\ref{sec:AM}, Eq.~(\ref{eq:corr})].
The fluctuations of the morphogen's concentration at two points separated by distances
greater than $\ell L = \sqrt{D/k}$ are nearly independent, whereas the decay of the
time correlations is controlled by $(k \ell^2)^{-1} = L^2/D$ (Appendix~\ref{sec:AM}).
The constants $k$ and $L$ determine the scale of the system; they can serve as units
of time and length.

\section{\label{sec:num}Numerical results}
As an application of the Van Kampen theory, we estimate the level of fluctuations
for the concentration of the morphogen bicoid in a {\it Drosophila} embryo \cite{Gregor2007Probing,Grimm2010}.
The results of our calculations are reported in a system of units reduced by the
length constant $L$ and the time constant $k^{-1}$. The values of the physical parameters
are adopted from experimental data \cite{Gregor2007Probing,Grimm2010}: $L=0.5$ mm,
$\lambda=0.1$ mm, $a_0=55$ nM (1 nM corresponds to 0.6 molecules/$\mu$m\textsuperscript{3}).
Converted to reduced units, the source strength and the correlation length are respectively
$a_0 = 4.125\times10^9$ and $\ell = 0.2$. The concentration of bicoid is presumably
read out by densely distributed cellular nuclei, whose spatial separation sets a
plausible coarse-grain scale of $\xi=0.02$.

Figure~\ref{fig:num}~(a) illustrates a typical dependence of the standard deviation
$\std(a_\xi)$ on the position $x$, calculated for the coarse-grained PI curve. This
is a convex curve that is defined over the interval $[\xi/2,L-\xi/2]$ and decreases
monotonically from its maximum at $x=\xi/2$. For large correlation lengths $\lambda = \ell L$,
the variance of the coarse-grained morphogen concentration depends almost linearly
on $x$ and flattens when $\lambda\to\infty$ and $k\to0$. In the latter case, which
represents pure diffusion without degradation, the fluctuations of $\alpha(x)$ are
maximal for any given values of $a_0>0$ and $\xi>0$. On the other hand, when the
rate constant $k$ becomes infinitely large, the problem degenerates and fluctuations
vanish.

Of the two numerical integration schemes discussed in this paper, the collocation
method (Appendix~\ref{sec:CM}) is less accurate than the spectral finite-element
algorithm (Appendix~\ref{sec:SM}). The latter approach compares favorably with the
analytical solution [Eq.~(\ref{eq:smn}) in Appendix~\ref{sec:AM}]. Both integration
algorithms nonetheless reproduce correctly the average concentration and the overall
shape of the variation profile.

The relative uncertainty of the PI curve in our numerical example does not exceed
$0.4\,\%$ at any position [Fig.~\ref{fig:num}~(b)]. Because $\alpha_\xi$ decreases
with $x$ faster than its standard deviation, the coefficient of variation increases
towards the right boundary. The relative uncertainty nevertheless remains within the
order of $0.2\,\%$ in most of the system. Even an error of three standard deviations
still yields a coefficient of variation within the order of $1\,\%$. The precision
of the PI readout might therefore be limited mainly by the efficiency of the morphogen's
receptors.

Because the modeled system is half a milimeter in length \cite{Gregor2007Probing},
the small uncertainty of the PI curve in our example comes as no surprise. Given
a target precision of $10\,\%$ \cite{Gregor2007Probing}, the concentration of bicoid
can be measured in a period of time that is very short in comparison to the correlation
scale $(k \ell)^{-1}$. This result validates the Van Kampen theory for conditions
approaching the macroscopic limit. However, in developmental processes on a scale
of tens of micrometers \cite{JacoboHudspeth2014}---the dimension relevant to the
specification of intracellular structures---fluctuations can challenge the efficiency
of morphogen receptors. Additional mechanisms, such as biochemical feedback loops
\cite{Dublanche2006}, might then be required to reduce noise in the system.

\section{Conclusion}
The Van Kampen theory provides a promising means of estimating the fluctuation level
in RD problems and more generally in systems of mesoscopic physical fields. The approach
is conceptually simple and has a relatively small computational cost. Although in
this paper we consider only the steady-state solution of a RD problem, transients
can be taken into account as well [Eq.~(\ref{eq:general}) in Appendix~\ref{sec:AM}].
Moreover, the Van Kampen theory can be integrated readily into multiscale computational
models.

To simulate the Van Kampen equation, we formulated and tested two numerical techniques.
The results of spectral finite-element simulations (Appendix~\ref{sec:SM}) were
quite accurate and superior to those of the collocation method (Appendix~\ref{sec:CM}).

As a case study we chose a relatively large, 500~$\mu$m-long system for its simple
geometry and the availability of experimental data. Because the length scale approaches
macroscopic conditions, fluctuations of the PI curve in our simple example are very
small. In many other instances of the PI problem, however, the system's size can
be 10~$\mu$m or even less. At such scales, the fluctuations of the PI curve can impose
operational time and space constraints on the detectors of morphogen concentration.
Estimation of the noise level might provide insight into the mechanisms of encoding
and readout of PI. For example, the concentration's uncertainty might help in identifying
a morphogen among the candidate substances that occur in a system.

In a study focused on a specific RD problem, there are more details that could be
included in a Van Kampen equation: fluctuations of the source strength, boundary
effects, and the dimensionality of the system. Incorporation of these factors should
improve the accuracy of a theoretical model (Appendices~\ref{sec:AM} and \ref{sec:bnd}).

\begin{acknowledgments}
The authors thank Dr. A. Erzberger, Dr. A. Milewski, and Dr. F. Berger for stimulating
discussions and valuable comments on our results.

A. J. is a Fellow of the F. M. Kirby Foundation. R. B. is a Research Associate and
A. J. H. an Investigator of Howard Hughes Medical Institute.
\end{acknowledgments}

\appendix
\section{\label{sec:SRD} Van Kampen reaction-diffusion equation}
The Van Kampen RD equation~(\ref{eq:SRD}) extends the Langevin model
of fluctuations for simple time-dependent physical quantities to spatially distributed
fields \cite{vanKampenXIV,vanKampen1976,Gardiner8}. Consider first the classical
RD dynamics for the number density $a(t,x)$ of some morphogen over the linear domain
$x\in\Lambda$:
\begin{equation}\label{eq:RD}
  \partial_t a(t,x) = - k a(t,x) + D\partial_x^2 a(t,x),
\end{equation}
in which the degradation rate $k$ and the diffusivity $D$ are positive constants. The
first term on the right-hand side of Eq.~(\ref{eq:RD}) states the mass-action law
for the chemical degradation of the morphogen. The second term represents the divergence
of the Fick's diffusion flow
\begin{equation}\label{eq:Fick}
  J(t,x) = -D\partial_x a \,\Rightarrow\, -\partial_x J(t,x) = D\partial_x^2 a,
\end{equation}
which describes the balance of incoming and outgoing currents of matter $J(t,x)$.

Both macroscopic laws---mass action and Fick's diffusion---emerge as statistical
averages of the microscopic dynamics \cite{vanKampenXIV,vanKampen1976} over a steady-state
ensemble. Ergodicity of a system is commonly assumed as well. At mesoscopic scales,
however, we must allow fluctuations by replacing the following terms in Eqs.~(\ref{eq:RD})
and (\ref{eq:Fick}):
\begin{align}\label{eq:rep}
  k a(t,x) &\to k a(t,x) - \chi(t,x),\nonumber\\
  J(t,x) &\to -D\partial_x a(t,x) - j(t,x),
\end{align}
in which $\chi(t,x)$ and $j(t,x)$ are the deviations from the classical macroscopic
laws of reaction and diffusion, respectively. Due to the spontaneous variations given
by Eqs.~(\ref{eq:rep}), the local change of concentration $\partial_t a(t,x)$ does
not exactly match the fluctuating force on the right-hand side of Eq.~(\ref{eq:RD}).
The residual is
\begin{equation}\label{eq:residual}
  (\partial_t + k - D\partial_x^2) a(t,x) = \chi(t,x) + \partial_x j(t,x).
\end{equation}
The spontaneous behavior of the fluctuating force on the right-hand side of this
equation appears practically random and is therefore modeled as a stochastic, spatially
distributed process.

In the Langevin approach, the macroscopic properties of a steady-state dynamics correspond
to the ensemble averages, here denoted by angle brackets, of mesoscopic variables.
In particular, Eq.~(\ref{eq:RD}) requires $\avg{\chi(t,x)} = 0$ and $\avg{\partial_x j(t,x)} =~0$.
Hence the ensemble average of Eq.~(\ref{eq:residual}) yields
\begin{equation}\label{eq:SS}
  k \avg{a(t,x)} - D\partial_x^2\avg{a(t,x)} = 0
\end{equation}
for $\avg{\partial_t a(t,x)} = 0$ by the definition of a steady state.

For a nontrivial solution $\avg{a(t,x)} \ne 0$ to exist, we model a source of the
chemical agent by a nonhomogeneous Dirichlet boundary condition, which fixes the
value of $\avg{a(t,x)}$ at $x = 0$:
\begin{equation}\label{eq:LNDirichlet} \avg{a(t,0)} = a_0. \end{equation}
A natural choice of the other boundary at $x=L$ is the homogeneous Neumann condition,
which reflects the diffusive flow $\avg{J(t,x)}$ [Eq.~(\ref{eq:rep})]:
\begin{equation}\label{eq:RHNeumann} \avg{\partial_x a(t,L)} = 0. \end{equation}
Subject to the above constraints, Eq.~(\ref{eq:SS}) is easy to solve \cite[Chapter 2]{EdwardsPenney}.
We thus obtain the macroscopic time-independent PI curve plotted in Fig.~\ref{fig:xmp},
\begin{equation}\label{eq:alpha}
  \alpha(x) = \avg{a(t,x)} = a_0 \frac{\cosh[(L-x)/\lambda]}{\cosh(L/\lambda)},
\end{equation}
in which $\lambda=\sqrt{D/k}$.

For Langevin dynamics~(\ref{eq:RHNeumann}) we reduce the homogeneous Neumann
condition at the right end to a consistent nonhomogeneous Dirichlet boundary:
\begin{equation}\label{eq:RNDirichlet} \avg{a(t,L)} = \alpha(L). \end{equation}
As discussed in Appendix~\ref{sec:bnd}, both Dirichlet conditions~(\ref{eq:LNDirichlet})
and (\ref{eq:RNDirichlet}) neglect fluctuation effects at the boundaries of the RD
system. These effects can be included by imposing the reflective Neumann conditions
on both ends of $\Lambda$. These details, which are not strictly necessary for a
simple demonstration of the Van Kampen theory, are spared for Appendix~\ref{sec:bnd}.

The Langevin model is complete once the statistical properties for the right-hand
side of Eq.~(\ref{eq:residual}) have been specified. Van Kampen derives them by reducing
the microscopic RD dynamics to a random walk, a traditional argument of statistical
mechanics \cite{[{Chapter I in }] []Chandrasekhar,*Piasecki}. A continuous limit
of this simplified model gives
\begin{align}
  \avg{\chi(t,x) \chi(\prm{t},\prm{x})} = k \alpha(x) \delta(t-\prm{t}) \delta(x-\prm{x}),\\
  \avg{j(t,x) j(\prm{t},\prm{x})} = 2 D \alpha(x) \delta(t-\prm{t}) \delta(x-\prm{x}),
\end{align}
which hold for any instants of time $t,\prm{t}$, and positions $x,\prm{x}$ \cite{vanKampenXIV,vanKampen1976}.
The theory behind the above equations relies on the following assumptions: $\chi$ and $j$ are independent
($\avg{\chi(t,x) j(\prm{t},\prm{x})} \equiv 0$); an infinitesimal interval $dx$ contains
a large number of molecules $\alpha(x) dx$; and {\it all} correlations at distances
of order $dx$ are negligible. Then infinitesimal processes $\chi(t,x)$ and $j(t,x)$
are approximately Gaussian by virtue of the central limit theorem \cite[Sec. 2.5]{KardarI}.

The Van Kampen model leads directly to the concept of a spatially distributed Gaussian
white noise $\partial_x \dot{W}(t,x)$ with a zero mean and a constant strength $\beta$.
The defining property of $\partial_x \dot{W}$ is that its integral over a time interval
$t$ and a line segment $\Xi(x)=(x-\xi/2,x+\xi/2)$,
\begin{equation}\label{eq:Wiener1}
  W(t,x|\xi) = \int_0^t \prm{dt} \int_{\Xi(x)} \prm{dx} \partial_{\prm{x}} \dot{W}(\prm{t},\prm{x}),
\end{equation}
is a Gaussian random process of zero mean ($\avg{W} = 0$) and variance
\begin{equation}\label{eq:Wiener2} \avg{W(t,x|\xi)^2} = \beta \xi t. \end{equation}
All properties of the stochastic processes $\chi(t,x)$ and $j(t,x)$ are then encompassed
by
\begin{align}\label{eq:noise1}
  \chi(t,x) = \sqrt{k \alpha(x)} \partial_x \dot{W}_1(t,x),
  \\\label{eq:noise2}
  j(t,x) = \sqrt{2 D \alpha(x)} \partial_x \dot{W}_2(t,x),
\end{align}
in which $\partial_x \dot{W}_1$ and $\partial_x \dot{W}_2$ are two independent, spatially
distributed, Gaussian white-noise terms of unit strength~$\beta=1$ [Eqs.~(\ref{eq:uncorrelated})
and (\ref{eq:white})]. Note that $j(t,x)$ is a vector quantity, which in one dimension
has a single component $\partial_x \dot{W}_2(t,x)$.

\section{\label{sec:AM} Green's function method}
Supplemented with an initial value $a(0,x)$ and the boundary conditions~(\ref{eq:bnd}),
Eqs.~(\ref{eq:residual})--(\ref{eq:noise2}) lead to Eq.~(\ref{eq:SRD}):
$$ (\partial_t + k - D\partial_x^2) a(t,x) = f(t,x). $$
This stochastic partial differential equation is linear, as is its left-hand-side
operator $\oper{L} = (\partial_t + k -D\partial_x^2)$, and inhomogeneous, in that
$f(t,x)$ enters the expression additively.

A general solution of Eq.~(\ref{eq:SRD}) is most conveniently expressed with the
help of the Green's function \cite[Chapter 10]{Arfken7} $g(t-\prm{t}, x|\prm{x})$,
which we find from the following equations:
\begin{align}\label{eq:GRD}
  &\oper{L} g(t-\prm{t}, x|\prm{x}) = \delta(t-\prm{t}) \delta(x-\prm{x}),\\
  \label{eq:gbnd}
  &g(t,0) = g(t,L) = 0.
\end{align}
In a finite domain $\Lambda$ the Green's function can be expanded as a series
\cite{[{Secs. 2.4 and 4.2 in }] []Duffy2001,*[{Secs. 3.4 and 5.2 in }] []Duffy2015}.
For the problem at hand we use a discrete expansion ($n=1,2...$) in the orthonormal
Fourier basis
\begin{equation}\label{eq:basis}
  \phi_n(x) = \sqrt{2/L} \sin(n \pi x/L),
\end{equation}
which is complete under the boundary conditions~(\ref{eq:gbnd}). One then finds
\begin{align}\label{eq:Green}
  &g(t-\prm{t},x|\prm{x}) = \sum_{n=1}^{\infty} g_n(t-\prm{t}) \phi_n(\prm{x}) \phi_n(x),
  \\\label{eq:gn}
  &g_n(t) = H(t) \exp\left\{
    -t\left[k+D\frac{n^2\pi^2}{L^2}\right]
  \right\},
\end{align}
in which $H(\cdot)$ stands for the Heaviside step function.

With the boundary conditions (\ref{eq:bnd}) and (\ref{eq:gbnd}), the general solution
of (\ref{eq:SRD}) takes the form
\begin{multline}\label{eq:general}
  a(t,x) = \alpha(x) + \int_\Lambda \prm{dx} g(t,x|\prm{x}) [a(0,\prm{x}) - \alpha(\prm{x})]\\
    + \int_0^t \prm{dt} \int_\Lambda \prm{dx} g(t-\prm{t},x|\prm{x}) f(\prm{t},\prm{x}),
\end{multline}

in which the second term on the right-hand side vanishes as $\lim_{t\to\infty} g(t,x|\prm{x})=0$.
If one is concerned merely with the steady-state behavior of Eq.~(\ref{eq:SRD}),
the transient solutions can be eliminated from Eq.~(\ref{eq:general}) by taking the
limit of infinite $t$:
\begin{multline}\label{eq:steady}
  a(\infty,x) = \alpha(x) \\+ \lim_{t\to\infty}\int_0^t \prm{dt} \int_\Lambda \prm{dx} g(t-\prm{t},x|\prm{x}) f(\prm{t},\prm{x}).
\end{multline}

Consider the statistical properties of the steady-state solution $a(\infty,x)$. Because
$f(t,x)$ given by Eqs.~(\ref{eq:ftx}) is a linear superposition of zero-mean, Gaussian
white-noise terms, the ensemble average of Eq.~(\ref{eq:steady}) is consistent with
the macroscopic dynamics~(\ref{eq:RD}):
$$\avg{a(\infty,x)} = \alpha(x).$$
The second cumulant $\kappa_2[a(\infty,x)]$ of $a(\infty,x)$ can be obtained from
Eqs.~(\ref{eq:bnd})--(\ref{eq:delta}), and (\ref{eq:gbnd}):
\begin{multline}\label{eq:variance}
  \kappa_2[a(\infty,x)] = \kappa_2[\Delta{a}(\infty,x)] = \avg{ a(\infty,x)^2 - \alpha(x)^2}
    \\= k \lim_{t\to\infty} \int_0^t \prm{dt} \int_\Lambda \prm{dx} \alpha(\prm{x}) [g(t-\prm{t},x|\prm{x})]^2
    \\+ 2D \lim_{t\to\infty} \int_0^t \prm{dt} \int_\Lambda \prm{dx} \alpha(\prm{x}) [\partial_{\prm{x}} g(t-\prm{t},x|\prm{x})]^2.
\end{multline}
Higher-order cumulants of the steady-state solution vanish due to the Gaussian nature
of $f(t,x)$ and hence of $\alpha(\infty,x)$ as well.

As explained in Sec.~\ref{sec:thr}, the formal expression~(\ref{eq:variance}) diverges.
Therefore, to estimate the magnitude of fluctuations in the morphogen's concentration,
we calculate the variance of the coarse-grained number density $a_\xi(\infty,x)$
from Eq.~(\ref{eq:kappa2}). This computation can be carried out through a series
expansion of the Green's function, $g(t-\prm{t},x|\prm{x})$, truncated at $N$th term.
Let us introduce the following formulas:
\begin{align}\label{eq:notation}
  &\Phi_n(x) = \int_{\Xi(x)} \frac{\prm{dx}}{\xi} \phi_n(\prm{x}) = \frac{2 L}{n \pi \xi} \sin\left(\frac{n \pi \xi}{2 L}\right)\phi_n(x);
  \nonumber\\
  &\Omega_{mn} = \frac{4 \pi^2 \ell^3 m n \tanh[\ell^{-1}]}{[1 + \pi^2\ell^2 (m - n)^2] [1 + \pi^2\ell^2 (m + n)^2]},
\end{align}
in which $\ell = \sqrt{D/k}L^{-1}$. Then, by substituting Eqs.~(\ref{eq:alpha}),
(\ref{eq:Green}), and (\ref{eq:gn}) into (\ref{eq:kappa2}) and completing the integrals,
we obtain
\begin{equation}\label{eq:smn}
  \kappa_2[a_\xi(x)] = a_0 \sum_{mn} \Omega_{mn} \Phi_m(x) \Phi_n(x),
\end{equation}
in which the summation runs over all positive integers $m$ and $n$ up to $N$ ($m,n=1..N$).
Note that the mean value of the coarse-grained field $a_\xi(x)$ is
\begin{equation}\label{eq:axi}
  \avg{a_\xi(x)} = \frac{2\lambda}{\xi}\sinh\left(\frac{\xi}{2\lambda}\right) \alpha(x)
    \underset{\xi\to0}{\to} \alpha(x).
\end{equation}

We can similarly obtain the autocorrelation function $\kappa_2[a_\xi(0,x_1),a_\xi(t,x_2)]$
for the time-dependent concentration
\begin{equation}
  a_\xi(t,x) = \int_{\Xi(x)} \prm{dx} a(t,\prm{x}) = \alpha_\xi(x) + \Delta{a}_\xi(t,x).
\end{equation}
In linear systems the decay of the temporal and spatial autocorrelations is encompassed
by the Green's function \cite[Sec. 8.6]{Chandler}:
\begin{multline}\label{eq:corr}
  \kappa_2[a_\xi(0,x_1),a_\xi(t,x_2)] = \avg{\Delta{a}_\xi(0,x_1)\Delta{a}_\xi(t,x_2)}
    = \\ \int_\Lambda \prm{dx} g(t,x_2|\prm{x}) \avg{\Delta{a}_\xi(0,x_1) \Delta{a}(0,\prm{x})}
    = \\ a_0 \sum_{mn} \Omega_{mn} \Phi_m(x_1) \Phi_n(x_2) \exp[-k t (1+\pi^2 \ell^2 m^2)],
\end{multline}
in which only the transient term of Eq.~(\ref{eq:general}) makes a non-zero contribution
\cite{PRE2016III}. From Eqs.~(\ref{eq:Green}), (\ref{eq:gn}), (\ref{eq:notation}),
and (\ref{eq:corr}) it follows that, in reduced units (Sec.~\ref{sec:num}), the time
and space correlations are controlled respectively by the parameters $(k\ell)^2 = D/L^2$
and $\lambda = \ell L = \sqrt{D/k}$ through the diffusion constant $D$.

In computations the series expansion~(\ref{eq:smn}) and (\ref{eq:corr}) should be
truncated at an order $N \ge \ceil{2 L/\xi}$. This optimal value is suggested by
the following argument. Suppose that $2L/\xi$ is an integer. The Fourier mode $\phi_{n+N}(\cdot)$
is then an alias of $\phi_{n}(\cdot)$, because $\phi_{n+N}(x)=\phi_n(x)$ whenever
$x$ is an integer multiple of $\xi$. In Eq.~(\ref{eq:smn}) we passed from the basis
set $\phi_n(\cdot)$ to the coarse-grained functions $\Phi_n(\cdot)$ by integrating
over a spatial scale $\xi$ [Eq.~(\ref{eq:notation})]. This procedure allows us to
disregard aliasing modes with $n > N$. Indeed, spatial features smaller than the
scale $\xi$ should be smoothed by the coarse-graining integration. The regions near
the ends of the domain $\Lambda$ ($x\approx\xi/2,L-\xi/2$) are exceptions that may
require more terms to reduce ringing artifacts.

\section{\label{sec:SM} Spectral method}
Modal analysis similar to that of Appendix~\ref{sec:AM} leads to a simple method
of spectral finite elements \cite{Trefethen,vanDeVosse,Boyd} for Eq.~(\ref{eq:SRD}).
Subject to the boundary conditions (\ref{eq:bnd}), the number density $a(t,x)$ has
a series representation in terms of the basis functions given by Eq.~(\ref{eq:basis}):
\begin{equation}\label{eq:expansion}
  a(t,x) = \alpha(x) + \sum_{m=1}^\infty a_m(t) \phi_m(x),
\end{equation}
in which the time-dependent coefficients $a_m(t)$ must vanish on average to satisfy
Eq.~(\ref{eq:avgs}): $\avg{a_m(t)} = 0$.

Spatially distributed white noise $\partial_x \dot{W}_i$ ($i=1,2$) likewise has a
representation
\begin{equation}\label{eq:noise}
  \partial_x\dot{W}_i(t,x) = \sum_{n=1}^{\infty} \dot{w}_{in}(t) \phi_n(x),
\end{equation}
in which each time-dependent coefficient $\dot{w}_{in}(t)$ is a simple, independent
Gaussian white noise. Equations~(\ref{eq:uncorrelated}), (\ref{eq:white}), (\ref{eq:Wiener1}),
and (\ref{eq:Wiener2}) readily follow from Eq.~(\ref{eq:noise}). In higher dimensions
there are additional vector components like $\partial_x \dot{W}_2$ (Appendix~\ref{sec:AM})
that are independent in an orthogonal reference frame and therefore can be expanded
separately in series (\ref{eq:noise}). In nonorthogonal coordinate systems one must
also account for correlations due to the overlap of basis vectors.

Equations~(\ref{eq:expansion}) and (\ref{eq:noise}), truncated at some $m$ and $n$,
provide the finite-element representations of $a(t,x)$ and the white-noise terms.
Their substitution into (\ref{eq:SRD}) and a few simple manipulations eventually
lead to a system of ordinary differential equations for the coefficients $a_m(t)$:
\begin{equation}\label{eq:dam}
  \dot{a}_m(t) = -k (1+\pi^2\ell^2 m^2) a_m(t) + f_m(t),
\end{equation}
in which $\ell = \sqrt{D/k}L^{-1}$, and
\begin{align}\label{eq:Fmt}
  &f_m(t) = \sum_n \left[
     F_{1mn} \dot{w}_{1n}(t) + F_{2mn}\dot{w}_{2n}(t)
  \right],
  \\\label{eq:F1mn}
  &F_{1mn} = \int_\Lambda dx \sqrt{k\alpha(x)} \phi_m(x) \phi_n(x),
  \\\label{eq:F2mn}
  &F_{2mn} = -\int_\Lambda dx \sqrt{2 D \alpha(x)} \partial_x \phi_m(x) \phi_n(x).
\end{align}
From Eq.~(\ref{eq:dam}) we see that stochastic forces $f_m(t)$ randomly perturb the
modal coefficients $a_m(t)$. When explicit analytical expressions are not available
for the spatial integrals in Eqs. (\ref{eq:F1mn}) and (\ref{eq:F2mn}), a discrete
Fourier transform can be used instead as an approximation. This approach should then
be termed a pseudospectral finite-element method.

For a general RD system, the derivation of equations analogous to (\ref{eq:dam})
is quite straightforward. Because the problem studied in this paper is relatively
simple, we can obtain each coefficient $a_m(t)$ explicitly (Appendix~\ref{sec:AM}).
Because numerical methods are more widely applicable than analytical ones, however,
we develop below a pseudospectral finite-element scheme to solve the system of equations~(\ref{eq:dam}).

The system of equations~(\ref{eq:dam}) can be numerically integrated in time by
various methods, such as the Crank-Nicolson algorithm discussed in the next section
of the Appendix. We can also use a second-order stochastic operator-splitting
technique \cite[Appendix C]{PRE2017}:
\begin{align}\label{eq:sim}
  a_m(t+\Delta{t}) =& \exp[-(1+\pi^2 \ell^2 m^2) k \Delta{t}] a_m(t)
  \nonumber\\
    +& \exp\left[-(1+\pi^2 \ell^2 m^2) \frac{k \Delta{t}}{2}\right] F_m(\Delta{t});
  \nonumber\\
  F_m(\Delta{t}) =&  \int_0^{\Delta{t}} dt\,f_m(t)\nonumber\\
    =& \sqrt{\Delta{t}} \sum_n [
      F_{1mn} w_{1n} + F_{2mn} w_{2n}
    ],
\end{align}
in which $w_{1n}$ and $w_{2n}$ are independent Gaussian random variables of zero
mean and unit variance, whereas $\Delta{t}$ is an integration time step.

By coarse-graining Eq.~(\ref{eq:expansion}), one easily finds a finite-element representation
of $\Delta{a}_\xi(x) = a_\xi(x) - \alpha_\xi(x)$ in the notation of Eq.~(\ref{eq:notation}):
\begin{equation}
  \Delta{a}_\xi(x) = \sum_m a_m \Phi_m(x),
\end{equation}
in which the coefficients $a_m$ are correlated Gaussian variables. Their covariance
matrix
$$K_{mn} = \avg{a_m a_n}$$
can be sampled in a numerical simulation of Eq.~(\ref{eq:sim}). The variance of $\Delta{a}_\xi(x)$
is then equal to
\begin{equation}
  \kappa_2[\Delta{a}_\xi(x)] = \sum_{mn} K_{mn} \Phi_m(x) \Phi_n(x).
\end{equation}
By comparing the above equation with (\ref{eq:smn}), we see that $K_{mn} = a_0 \Omega_{mn}$.

The above results show that the modal coefficients $a_m(t)$ are correlated Gaussian
random variables of finite mean and variance. However, there are so many of them
in the representation of $a(\infty,x)$ that its variance given by Eq.~(\ref{eq:variance})
diverges unless the coarse-grained basis functions $\Phi_m(x)$ are used as in (\ref{eq:smn}).
Thanks to the nonsingular nature of the modal coefficients, a computer simulation
of Eq.~(\ref{eq:sim}) is feasible.

The series expansions (\ref{eq:expansion}) and (\ref{eq:noise}) need not have the same
number of modes. The argument of Appendix~\ref{sec:AM} for Eq.~(\ref{eq:smn}) sets
the optimum for $m = 1,2,..N_m$, in which $N_m = \ceil{2 L/\xi}$. The accuracy of
simulations can be improved indefinitely by taking progressively more terms in Eq.~(\ref{eq:noise}),
$N_n \ge N_m$. In Sec.~\ref{sec:num} we report our results for $N_n = 4 N_m$ and
$\Delta{t}=10^{-3}$.

\section{\label{sec:CM} Collocation method}
The Fourier components $a_m(t)$ of the morphogen's concentration are Gaussian random
variables of finite variance (Appendix~\ref{sec:SM}). This vector representation
of the field $a(t,x)$ can be projected onto another basis set. Then, in principle,
it should be possible to simulate numerically the dynamics of the new components.

\begin{figure}
\includegraphics[width=1\columnwidth]{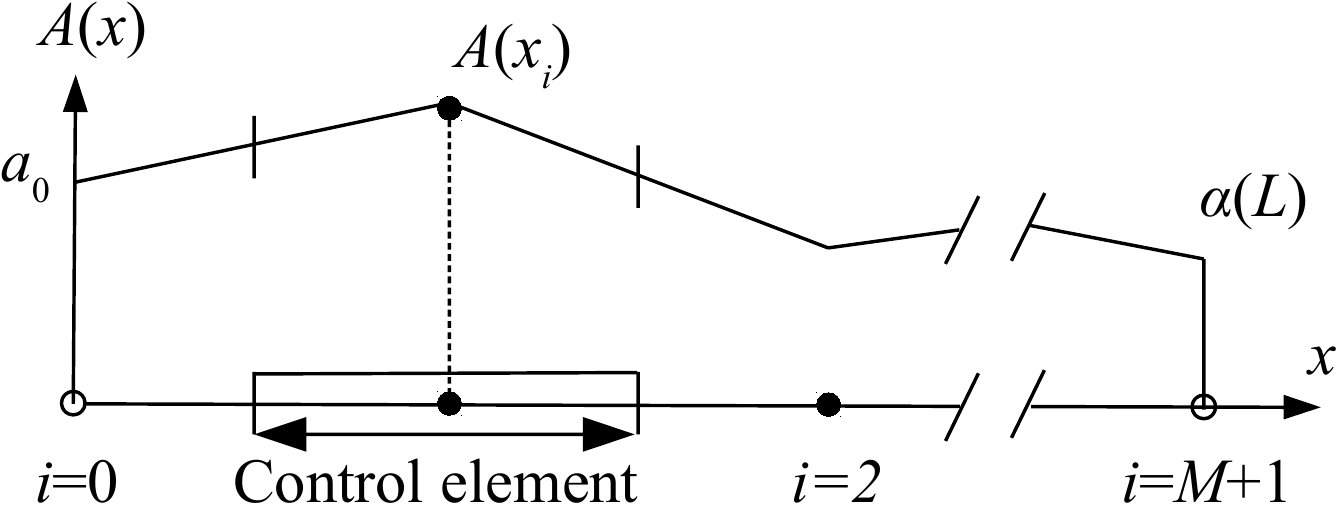}
\caption{\label{fig:fvm}
  A piecewise-linear interpolant $A(x)$ on a grid $x_i,\,i=0..M+1$. We use the boundary
  conditions (\ref{eq:bnd}) to fix the values $A(x_0)=a_0$ and $A(x_{M+1}) = \alpha(L)$
  at the nodes $i=0$ and $i=M+1$ ($\circ$), respectively. The control elements are
  centered on the nodes $i=1..M$ ($\bullet$). This collocation scheme reserves small
  intervals near the boundaries of the domain for a convenient implementation of
  the coarse-graining procedure.
}
\end{figure}

In this section we use a piecewise-linear interpolation \cite[Chapter 1]{Trefethen},
as a basis set for the finite-volume method, a widely used collocation finite-element
scheme \cite[Chapter 4]{Mattiussi1997,FVM}. Consider a uniform grid $x_i = i\Delta{x},\, i=0,1,\dots M+1$
on the domain $\Lambda$ (Fig.~\ref{fig:fvm}). We center control elements of size
$\Delta{x}$ at the nodes $x_i,\,i=1,2,\dots M$. The morphogen's concentration is
then interpolated by
\begin{eqnarray}\label{eq:interpolant}
  a(t,x) &\approx& \sum_{i=0}^{M+1} A_i(t) \eta_i(x),\\
  \eta_i(x) &=& \begin{cases}
    \frac{\Delta{x}-x_i+x}{\Delta{x}},\;\text{if } x_{i-1} \le x \le x_i\\
    \frac{\Delta{x}+x_i-x}{\Delta{x}},\;\text{if } x_i \le x \le x_{i+1}\\
  \end{cases},
\end{eqnarray}
which coincides with $a(t,x)$ at the centers of the control elements $A_i(t) = a(t,x_i)$.
To comply with the boundary conditions~(\ref{eq:bnd}), we fix $A_0 = a_0$ and $A_{M+1} = \alpha(L)$.
Thus the above interpolation is completely specified by $M$ time-dependent components
$A_i(t),\,i=1..M$.

A standard procedure of the finite-volume method would be to integrate Eq.~(\ref{eq:SRD})
over each $i$th control element $X_i = [x_i-\Delta{x}/2,x_i+\Delta{x}/2]$. In addition
to this, we must apply coarse graining over the scale $\xi \ge \Delta{x}$ in order
to remove the spatial singularity of the stochastic noise (Sec.~\ref{sec:thr}). It
is convenient to partition the domain $\Lambda$ so that $\xi$ is an integer multiple
of $\Delta{x}$: $\xi = P\Delta{x}$. We construct an integral operator
\begin{equation}\label{eq:Pint}
  \oper{I}_{Pi} = P^{-1}\sum_{j=0}^{P-1} \oper{I}_{i+j} = P^{-1}\sum_{j=0}^{P-1} \iint_{X_{i+j}} \frac{dx^2}{\Delta{x}}.
\end{equation}
In the above expression it suffices to consider only $P=1$ with a single term $\oper{I}_i$
that can be used to evaluate Eq.~(\ref{eq:Pint}) for an arbitrary $P$. Applied to
$a(t,x)$, the operator $\oper{I}_i$ gives a spatial integral of the coarse-grained
concentration:
\begin{multline}
  \oper{I}_i a(t,x) = \int_{X_i} dx_1
    \int_{x_1-\Delta{x}/2}^{x_1+\Delta{x}/2}\frac{dx_2}{\Delta{x}}\,a(t,x_2)
    \\= \int_{X_i} dx_1\, a_\xi(t,x_1)\Big{|}_{\xi=\Delta{x}}.
\end{multline}

By applying each of the $M$ operators $\oper{I}_i,\,i=1,2 \dots M$ to both sides of Eq.~(\ref{eq:SRD}),
we get
\begin{multline}
    (\partial_t + k) \oper{I}_i a(t,x)
      \\- \frac{D}{\Delta{x}} [a(t,x_i-\Delta{x}) - 2 a(t,x_i) + a(t,x_i+\Delta{x})]
      \\= \oper{I}_i f(t,x).
\end{multline}
Then substituting Eq.~(\ref{eq:interpolant}) for $a(t,x)$ yields
\begin{multline}\label{eq:dotA}
    \Delta{x} (\partial_t + k) \left[\frac{2 A_i(t)}{3} + \frac{A_{i-1}(t) + A_{i+1}(t)}{6}\right]
      \\- \frac{D}{\Delta{x}} [A_{i-1}(t) - 2 A_i(t) + A_{i+1}(t)]
      \\= \oper{I}_i f(t,x),
\end{multline}
which forms a system of $M$ ordinary differential equations to be solved for $A_i(t),i=1,2\dots M$.

The statistical properties of the coarse-grained field $a_\xi(t,x)$ can be estimated
from the values $A_i(t)$ [Eq.~(\ref{eq:interpolant})] sampled in a computer simulation.
For example, when $\xi=\Delta{x}$ one obtains
\begin{eqnarray}\label{eq:Axi}
  a_{\Delta{x}}(t,x_i) &\approx& \frac{3 A_i(t)}{4} + \frac{A_{i-1}(t) + A_{i+1}(t)}{8},
  \nonumber\\
  a_{\Delta{x}}(t,x_i + \Delta{x}/2) &\approx& \frac{A_i(t)}{2} + \frac{A_{i+1}(t)}{2}.
\end{eqnarray}
To integrate Eq.~(\ref{eq:dotA}) in time we use the Crank-Nicolson scheme based on
the trapezoidal rule
\begin{equation}\label{eq:tint}
  \int_t^{t+\Delta{t}} dt A_i(t) \approx [A_i(t) + A_i(t+\Delta{t})]\frac{\Delta{t}}{2}.
\end{equation}
The simulation algorithm can be concisely written in the vector-matrix notation
\begin{eqnarray}\label{eq:fvm}
  \mtx{T} \mtx{A}(t+\Delta{t}) &=& \mtx{E}\mtx{A}(t) + \mtx{R}(\Delta{t});\\
  T_{ii} &=& (2+k\Delta{t})\frac{\Delta{x}}{3} + D\frac{\Delta{t}}{\Delta{x}},\\
  T_{i(i\pm1)} &=& (2+k\Delta{t})\frac{\Delta{x}}{12} -D\frac{\Delta{t}}{2\Delta{x}},\\
  E_{ii} &=& (2-k\Delta{t})\frac{\Delta{x}}{3} - D \frac{\Delta{t}}{\Delta{x}},\\
  E_{i(i\pm1)} &=& (2-k\Delta{t})\frac{\Delta{x}}{12} + D\frac{\Delta{t}}{2\Delta{x}},
\end{eqnarray}
with all other elements of $M$-by-$M$ matrices $\mtx{T}$ and $\mtx{E}$ being zero.
The column vector $\mtx{R}(\Delta{t})$ of size $M$ is given below.

On the right-hand side of Eq.~(\ref{eq:dotA}) we obtain two contributions, which correspond to
the stochastic terms of the force $f(t,x)$ [Eq.~(\ref{eq:ftx})]:
\begin{align}\label{eq:uv}
  &\int_{\Delta{t}} dt \oper{I}_i \sqrt{k \alpha(x)} \partial_x \dot{W}_1(t,x) = u_i,\nonumber\\
  &\int_{\Delta{t}} dt \oper{I}_i \partial_x [\sqrt{2 D \alpha(x)} \partial_x \dot{W}_2(t,x)]
    = v_i - v_{i+1},
\end{align}
in which $u_i$ and $v_i$ are Gaussian random variables of zero mean. The covariances
of $u_i$ and $v_i$ can be calculated from Eq.~(\ref{eq:uncorrelated}) and (\ref{eq:white}).
In particular, all $v_i$ are independent of each other, as well as from $u_i$ which
correlate in pairs: $\avg{u_i u_j}\ne0$ if $|i-j|=1$. We omit lengthy calculations
that eventually lead to
\begin{align}\label{eq:k2ui}
  \kappa_2(u_i) =& \frac{4 \lambda^3 k \Delta{t}\alpha(x_i)}{\Delta{x}^2} \left[
    \sinh\left( \frac{\Delta{x}}{\lambda} \right) - \frac{\Delta{x}}{\lambda}
  \right],
  \\\label{eq:k2vv}
  \avg{u_i u_{i+1}} &= \frac{4 \lambda^3 k \Delta{t} a_0}{\Delta{x}^2\cosh(L/\lambda)}
  \cosh\left(\frac{2 L - x_i - x_{i+1}}{2\lambda}\right)
  \nonumber\\ &\times\left[
    \frac{\Delta{x}}{2 \lambda}\cosh\left( \frac{\Delta{x}}{2\lambda} \right)
    -\sinh\left( \frac{\Delta{x}}{2\lambda} \right)
  \right],
  \\\label{eq:k2vi}
  \kappa_2(v_i) =& \frac{2 \lambda D \Delta{t} a_0}{\Delta{x}^2\cosh(L/\lambda)}
  \nonumber\\ &\times\left[
    \sinh\left( \frac{L-x_{i-1}}{\lambda} \right)
    -\sinh\left( \frac{L-x_i}{\lambda} \right)
  \right],
\end{align}
with $\lambda = \sqrt{D/k}$.

An alternative approach can be used for problems in which the derivation of formulas
analogous to Eqs.~(\ref{eq:k2ui})--(\ref{eq:k2vv}) is too tedious. If we substitute
the spectral representation of white noise~(\ref{eq:noise}) into Eq.~(\ref{eq:uv}),
we get
\begin{align}\label{eq:altuv}
  &u_i = \sqrt{\Delta{t}} \sum_j w_{1j} u_{ij},\quad v_i = \sqrt{\Delta{t}} \sum_j w_{2j} v_{ij},\nonumber\\
  &u_{ij} = \oper{I}_i [\sqrt{k \alpha(x)} \phi_j(x)],\nonumber\\
  &v_{ij} = \int_{X_i}\frac{dx}{\Delta{x}} \sqrt{2 D \alpha(x+\Delta{x}/2)} \phi_j(x+\Delta{x}/2),
\end{align}
in which $w_{1j}$ and $w_{2j}$ are independent Gaussian random variables with zero
mean and unit variance. The elements $u_{ij}$ and $v_{ij}$ can be evaluated by the
pseudospectral method (Appendix~\ref{sec:SM}). In essence, the above equations are
projections of white noise in a spectral representation onto the linear-interpolation
basis functions, a procedure mentioned in the beginning of this section.

Finally, we can express the column vector $\mtx{R}$ as
\begin{align}
  &\mtx{R} = \mtx{b} + \mtx{u} + \mtx{C} \mtx{v},\\
  &b_0 = a_0 \Delta{t}\left(\frac{D}{\Delta{x}} - \frac{k\Delta{x}}{6}\right),\\
  &b_M = \alpha(L) \Delta{t}\left(\frac{D}{\Delta{x}} - \frac{k\Delta{x}}{6}\right) -v_{M+1}\\
  &C_{ii} = 1,\quad C_{i(i+1)} = -1,
\end{align}
in which the vector $\mtx{b}$ incorporates the boundary terms; the components $b_{ij}$
and the matrix elements $C_{ij}$ that equal zero are not indicated.

The results reported in Sec.~\ref{sec:num} are obtained by using Eqs.~(\ref{eq:k2ui})--(\ref{eq:k2vi})
with $\Delta{t}=10^{-3}$ and $\Delta{x} = 2\times10^{-4}$.

\section{\label{sec:bnd} Fluctuation effects at the source and boundaries}
The Dirichlet conditions~(\ref{eq:bnd}) fix the value of $a(t,x)$ at the ends of
the domain $\Lambda$. The resulting solution of Eq.~(\ref{eq:SRD}) therefore neglects
fluctuation effects at the boundaries and, in particular, at the source of the morphogen.
Indeed, the Green's function we obtained in Appendix~\ref{sec:AM} is insusceptible
to any forces at the ends of the domain $\Lambda$, where it vanishes due to Eq.~(\ref{eq:gbnd}).
Note that the left Dirichlet boundary in Eq.~(\ref{eq:bnd}) is an implicit source
of the morphogen.

Assuming a closed RD system, in which matter does not leak through the ends of the
domain $\Lambda$, we replace the reflective Neumann boundary conditions (\ref{eq:bnd})
by
\begin{equation}\label{eq:alt}
   \partial_x a(t,x)\Big{|}_{x=0} = 0,\quad \partial_x a(t,x)\Big{|}_{x=L} = 0,
\end{equation}
whereby the macroscopic diffusion flux through the points $x=0,L$ vanishes [Eq.~(\ref{eq:Fick})].
Nonetheless we shall take special care that the fluctuations of the matter flow $j(t,x)$
in Eq.~(\ref{eq:rep}) do not violate the closed-system constraint.

Together with Eq.~(\ref{eq:alt}), we must model explicitly the source of the morphogen
$s(x)$. Given that this substance is generated from a densely concentrated substrate
at an effective rate $k_+$, one can pose
\begin{equation}\label{eq:source}
  s(x) = k_+ \delta(x) + \sqrt{k_+} \dot{w}_+(t) \delta(x),
\end{equation}
in which the second term, with simple white noise coefficient $\dot{w}_+(t)$, introduces
fluctuations of the source strength.

The new boundary conditions~(\ref{eq:alt}) require a change of the basis set $\phi_n(\cdot)\to\psi_n(\cdot)$
in the series expansion for the Green's function [Eq.~(\ref{eq:Green})], for the
concentration $a(t,x)$ [Eq.~(\ref{eq:expansion})], and finally for white noise $\partial_x \dot{W}_1$,
but not for $\partial_x \dot{W}_2$. The Neumann basis set has an additional mode
for $n=0$:
\begin{equation}\label{eq:omega}
  \psi_0(x) = L^{-1/2},\;\psi_n(x)= \sqrt{2/L} \cos(n \pi x / L).
\end{equation}
The Neumann boundary conditions also alter the expression for the PI curve
$\alpha(x)\to\nu(x)$:
\begin{equation}
  \nu(x) 
  = \frac{k_+}{k} \sum_{n=0}^{\infty} \frac{\psi_n(0)\psi_n(x)}{1+\pi^2\lambda^2 n^2/L}.
\end{equation}

White noise in the expression for the fluctuating diffusive flow, $\partial_x \dot{W}_2(t,x)$,
should be treated separately. Because this term models variations of matter flux
through a point $x$, its spatial derivative at $x=0,L$ is not well defined. With
the no-leak conditions, matter is not allowed to flow {\it through} the boundaries
of the domain $\Lambda$. To satisfy this requirement we must use the expansion Eq.~(\ref{eq:noise})
for $\partial_x \dot{W}_2(t,x)$.

If we were to enforce the expansion of $\partial_x {W}_2(t,x)$ in the basis set~(\ref{eq:omega})
instead of~(\ref{eq:basis}), through integration by parts in Eq.~(\ref{eq:delta}) we
would obtain a boundary term of the form
\begin{equation}
  \lim_{t\to\infty}\int_0^t \prm{dt}
  g(t-\prm{t},x|\prm{x}) \sqrt{2D \alpha(\prm{x})}
  \partial_{\prm{x}} {W}_2(\prm{t},\prm{x}) \Big{|}_{\prm{x}=0}^{L}.
\end{equation}
As for Eq.~(\ref{eq:divergent}), this expression has a spatial singularity because
the Green's function does not vanish at the Neumann boundaries. We cannot remedy
this problem by coarse graining, for we cannot integrate over a $\xi$-neighborhood
of the end points $x=0,L$.

If the morphogen is allowed to leak through the boundaries of the system, new point
sources of fluctations, similar to the second term in Eq.~(\ref{eq:source}), may
emerge. The Van Kampen theory should then be extended to include this contribution.

In summary, we introduced above three point sources of fluctuations: noise in the
morphogen's synthesis at $x=0$ and in its degradation at $x=0$ and $x=L$. In our
simplified model of Sec.~\ref{sec:thr}, however, there is a continuum of noise over
the interval $x\in(0,L)$. Provided the average total number of molecules in the system
is large, we do not expect that the addition of three isolated points, as suggested
in this section, would alter the level of fluctuations.

Systematically applying the above changes to the theory presented earlier, one can
derive results that incorporate fluctuations at the source of the chemical agent
and at the boundaries of the RD system. From the analysis of this section it also
follows that the Dirichlet-Neumann conditions~(\ref{eq:LNDirichlet}) and (\ref{eq:RHNeumann})
would not provide a full account of these phenomena. A more reasonable choice would
be either to increase the level of details by Eqs.~(\ref{eq:alt}) and (\ref{eq:source})
or to neglect the boundary effects altogether by using Eq.~(\ref{eq:bnd}).

\bibliographystyle{apsrev4-1}
\bibliography{master}
\end{document}